\documentclass[aps,pra,preprint,groupedaddress]{revtex4}
\usepackage{amsmath}
\usepackage{graphicx}
\begin{document}
	
%\preprint{}
	
\title{From Poynting vector to new degree of freedom of polarization}

	
\author{Xiao-Lu You and Chun-Fang Li}
\email[]{youxiaolu123@shu.edu.cn (XLY) and cfli@shu.edu.cn (CFL)}
%\thanks{}
\affiliation{Department of Physics, Shanghai University, 99 Shangda Road, 200444 Shanghai, China}
	
	
%\date{\today}
	
\begin{abstract}
	
Up till now, the Jones vector is, strictly speaking, only a notion about the state of polarization of plane electromagnetic waves though it is generally applied to paraxial fields approximately.
Here we generalize it to non-paraxial fields. The same as the Jones vector for plane waves, the generalized Jones vector for non-paraxial fields is global in the sense that it does not depend on the field position. This is achieved by investigating the effect of the polarization on the Poynting vector in the non-paraxial superposition of four plane waves.
Even more importantly, by doing so we find that in addition to the Jones vector, another degree of freedom, called the Stratton vector, is also needed to completely describe the state of polarization of non-paraxial fields. It is shown that the polarization described by the global Jones vector is dependent on the position. The position dependence of the polarization originates in the position dependence of the polarization bases. The Stratton vector specifies the way in which the polarization bases depend on the position. A general expression for the dependence of the Poynting vector on the Stratton and Jones vectors is also given.
	
\end{abstract}
	
%\keywords{}
	
\maketitle

\section{\label{Introduction} Introduction}

The Poynting vector is the intensity of energy flow in the electromagnetic field \cite{Stra}, a position-dependent quantity. The cycle-averaged value of the Poynting vector in a monochromatic field takes the form,
\begin{equation}\label{PoyV}
\mathbf{g}=\frac{1}{2} \Re(\mathbf{E} \times \mathbf{H}^*),
\end{equation}
where $\mathbf E$ and $\mathbf H$ are the electric and magnetic fields, respectively. About two decades ago, Katsenelenbaum \cite{Kats} studied the direction of the Poynting vector in an electromagnetic field that is formed by the superposition of four plane monochromatic waves of linear polarization. The four plane waves propagate in the directions that make the same acute angle with the $z$ axis. He found that in the transverse plane, the Poynting vector in some domains is oriented in the opposite direction to the propagation of the plane waves, especially when the constituent plane waves are highly non-paraxial. In other words, the $z$-component of the Poynting vector in those domains is negative. Such a phenomenon was recently utilized \cite{Kotl-SNKP} to explain the reverse energy flow in a sharp focus.

However, our investigations show that the phenomenon found by Katsenelenbaum lies with the choice of the concrete state of linear polarization of the constituent plane waves. When the state of linear polarization is appropriately adjusted, the $z$-component of the resultant Poynting vector cannot be negative. In order to investigate the effect of the polarization on the Poynting vector in the non-paraxial superposition, we generalize the notion of Jones vector of plane waves to non-paraxial fields. We find that in addition to the generalized Jones vector, another degree of freedom is also needed to completely describe the state of polarization of non-paraxial fields. The purpose of this paper is to make use of the four-plane-wave model to show how the Poynting vector reveals the role of the new degree of freedom in describing the state of polarization of non-paraxial fields. What is important to realize is that the conclusions that we draw from the discussions about the superposition of four plane waves can readily be extended to a non-paraxial monochromatic field of continuously-distributed angular spectrum. 

According to Ref. \cite{Kats}, we consider the following electric field of the non-paraxial superposition of four plane waves in free space,
\begin{equation}\label{E}
   \mathbf{E} (\mathbf{x})
  =\frac{1}{\sqrt{2 \varepsilon_0}} \sum_{i=1}^{4} \mathbf{a}_i \exp(i \mathbf{k}_i \cdot \mathbf{x}),
\end{equation}
where the complex-valued unit vectors $\mathbf{a}_i$ are the polarization vectors of the plane waves,
\begin{equation*}
\begin{split}
\mathbf{k}_1 &= k (\hat{z} \cos \vartheta +\hat{x} \sin \vartheta),  \\
\mathbf{k}_2 &= k (\hat{z} \cos \vartheta +\hat{y} \sin \vartheta),  \\
\mathbf{k}_3 &= k (\hat{z} \cos \vartheta -\hat{x} \sin \vartheta),  \\
\mathbf{k}_4 &= k (\hat{z} \cos \vartheta -\hat{y} \sin \vartheta),
\end{split}
\end{equation*}
are their wavevectors, $k$ is the wavenumber, $\hat{x}$, $\hat{y}$, and $\hat{z}$ denote the unit vectors along the corresponding axes, $\vartheta$ is the acute angle that the wavevectors make with the $z$ axis, the factor $\frac{1}{\sqrt{2 \varepsilon_0}}$ is introduced for convenience, and the time dependence is assumed to be $\exp(-i \omega t)$.
As is known, the two components of the Jones vector of a plane wave are the projections of its polarization vector onto a pair of polarization bases, which are orthogonal to its propagation direction. So in order to define the Jones vector for the polarization of a plane wave, it is essential to specify the polarization bases. If the plane wave is assumed to propagate along the $z$ axis, one usually takes the unit vectors along the $x$ and $y$ axes as the polarization bases \cite{Dama}. But here all the four wavevectors in Eq. (\ref{E}) point in different directions. None of them is along the $z$ axis. The usual practice is no longer applicable. Even so, for each constituent plane wave, one is always possible to choose a pair of mutually perpendicular unit vectors orthogonal to its wavevector as the polarization bases.
Denoting by $\mathbf{u}_i$ and $\mathbf{v}_i$ the polarization bases for each plane wave, which satisfy
\begin{equation}\label{triad}
    \mathbf{u}_i \cdot  \mathbf{v}_i =0, \quad
    \mathbf{u}_i \times \mathbf{v}_i =\frac{\mathbf{k}_i}{k},
\end{equation}
one expands the corresponding polarization vector as follows,
\begin{equation}\label{a:alpha}
    \mathbf{a}_i= \alpha_1 \mathbf{u}_i +\alpha_2 \mathbf{v}_i .
\end{equation}
The expansion coefficients $\alpha_1$ and $\alpha_2$ make up its Jones vector,
$\alpha \equiv \bigg(\begin{array}{c}
                        \alpha_1 \\ \alpha_2
                     \end{array}
               \bigg)$,
which satisfies the normalization condition
$\alpha^\dag \alpha =1$.
But the unit vectors $\mathbf{u}_i$ and $\mathbf{v}_i$ satisfying Eqs. (\ref{triad}) are arbitrary to the extent that a rotation about the associated wavevector $\mathbf{k}_i$ can be
performed. By this it is meant that the Jones vector $\alpha$ in expression (\ref{a:alpha}) for the polarization vector $\mathbf{a}_i$ must be defined with respect to some polarization bases. On the other hand, as mentioned above, the polarization bases $\mathbf{u}_i$ and $\mathbf{v}_i$ must be orthogonal to the associated wavevector $\mathbf{k}_i$. In other words, they must depend on the wavevector. Therefore, they cannot be the same for different constituent plane waves. A convenient and consistent way \cite{Li08} to specify the polarization bases for all the plane waves is to make use of the constant unit vector that was first introduced by Stratton \cite{Stra} and later by others \cite{Gree-W, Patt-A, Davi-P} in the representation of electromagnetic fields, denoted by $\mathbf I$, in the following way,
\begin{equation}\label{PB}
    \mathbf{u}_i =\mathbf{v}_i \times \frac{\mathbf{k}_i}{k},             \quad
    \mathbf{v}_i =\frac{\mathbf{I} \times \mathbf{k}_i}
                      {|\mathbf{I} \times \mathbf{k}_i |}.
\end{equation}
However, as the expansion coefficients in Eq. (\ref{a:alpha}), the two elements of the Jones vector do not necessarily depend on the wavevector. If this is the case, Eq. (\ref{E}) can be rewritten as follows,
\begin{equation}\label{E:UV}
\mathbf{E}(\mathbf{x})=\alpha_1 \mathbf{U} +\alpha_2 \mathbf{V},
\end{equation}
where
\begin{equation}\label{UV}
\mathbf{U}=\frac{1}{\sqrt{2 \varepsilon_0}}\sum_{i=1}^{4} \mathbf{u}_i \exp(i \mathbf{k}_i \cdot \mathbf{x}), \quad
\mathbf{V}=\frac{1}{\sqrt{2 \varepsilon_0}}\sum_{i=1}^{4} \mathbf{v}_i \exp(i \mathbf{k}_i \cdot \mathbf{x}).
\end{equation}
It is the common Jones vector $\alpha$ of the constituent plane waves that is argued to be the generalized Jones vector for the entire superposition field, which is defined with respect to the polarization bases $\mathbf U$ and $\mathbf V$. It is global in the sense that it does not depend on the position $\mathbf x$ in the field. This is the non-paraxial field that we are exclusively concerned with in the present paper. 
What is noteworthy is that the polarization bases in (\ref{UV}) are specified by the constant vector $\mathbf I$ via Eqs. (\ref{PB}). So the generalized Jones vector for a non-paraxial field must be defined with respect to some polarization bases in much the same way as the Jones vector for a plane wave. 
As a result, given the constant vector $\mathbf I$ in (\ref{PB}), different Jones vectors will mean different states of polarization via Eq. (\ref{E:UV}). On the other hand, 
given the Jones vector $\alpha$, the polarization of the non-paraxial field (\ref{E:UV}) will depend on the choice of the constant vector $\mathbf I$. It can thus be concluded that the constant vector $\mathbf I$ shows up as a degree of freedom that combines with the Jones vector to describe the polarization of non-paraxial fields.
For this reason, we will refer to the constant vector $\mathbf I$ as Stratton vector.

Nevertheless, being different from the polarization bases $\mathbf{u}_i$ and $\mathbf{v}_i$ for the constituent plane wave, which have nothing to do with the position $\mathbf x$, the polarization bases $\mathbf U$ and $\mathbf V$ in (\ref{UV}) vary with the position. Therefore, even though the Jones vector is global, the polarization that it describes together with the Stratton vector is dependent on the position. Specifying the way in which the polarization bases vary with the position through Eqs. (\ref{PB}) and (\ref{UV}), the Stratton vector determines how the polarization described by the global Jones vector depends on the position. In a word, the polarization of the non-paraxial superposition of the four plane waves depends on the position even though the generalized Jones vector to describe it does not.
So if we refer to the position-dependent properties of an electromagnetic field as its locality properties, the polarization is such a property. To our satisfaction, the Poyning vector as another locality property is found to aptly reflect the above-mentioned features of the polarization. It is simply related to the Stratton and Jones vectors through the polarization bases and the polarization ellipticity, respectively. This is why the study of the Poynting vector makes it possible for us to discover that the Stratton vector is a new degree of freedom of the polarization. The contents of this paper are arranged as follows.

Following Katsenelenbaum, we construct in Section \ref{Examples} two different kinds of superposition fields to illustrate the dependence of the Poynting vector on the state of polarization. Each of them consists of four linearly-polarized monochromatic plane waves. One is similar to what Katsenelenbaum discussed in the sense that the $z$-component of its Poynting vector in some domains is negative. The other does not show negative $z$-component at all in the Poynting vector.
Such a difference is characterized in Section \ref{Explanation} with the help of the Stratton vector as well as of the common Jones vector of the constituent plane waves. It is shown that even though the common Jones vectors in both situations are the same, the polarization bases to define the common Jones vector are specified by different Stratton vectors. The dependence of the Poynting vector on the common Jones vector, when extended to include any non-zero ellipticity of polarization, is also discussed in each situation.
In Section \ref{Generalization}, the notion of Jones vector for plane waves is generalized to non-paraxial fields. It is argued that the common Jones vector of the constituent plane waves is the Jones vector of the entire superposition field (\ref{E:UV}). The roles of the generalized Jones vector and of the Stratton vector in describing the state of polarization of the superposition field are analyzed. A general expression to show the effects of the Stratton and Jones vectors on the Poynting vector is given. Section \ref{Conclusions} summarizes the paper with remarks. It is particularly pointed out how the superposition field considered by Katsenelenbaum
can be described by the Stratton and Jones vectors.

\section{\label{Examples} Dependence of Poynting vector on state of polarization}

\begin{figure}[t]
	\centerline{\includegraphics[width=6.5cm]{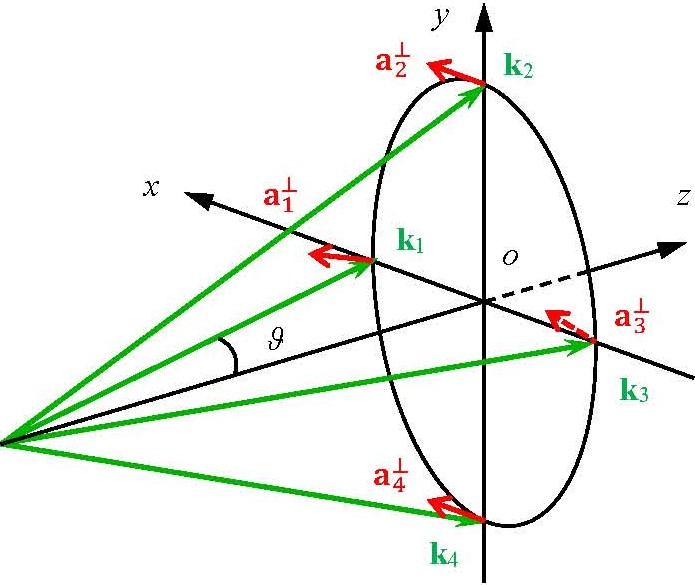}}
	\caption{\label{wavevector} Illustration for the polarization vectors of the four plane waves in the first situation.}
\end{figure}

Let us first consider a situation that is similar to what Katsenelenbaum discussed. The polarization vectors of the electric fields of the four constituent plane waves are chosen as follows,
\begin{equation}\label{PV}
\begin{split}
\mathbf{a}^{\perp}_1 &= \hat{x} \cos \vartheta -\hat{z} \sin \vartheta,  \\
\mathbf{a}^{\perp}_2 &= \hat{x},                                         \\
\mathbf{a}^{\perp}_3 &= \hat{x} \cos \vartheta +\hat{z} \sin \vartheta,  \\
\mathbf{a}^{\perp}_4 &= \hat{x},
\end{split}
\end{equation}
where the meaning of the superscript $\perp$ will be clear in the next section. They are all linearly-polarized and are schematically indicated in Fig. \ref{wavevector} by the red arrows. Substituting expressions (\ref{PV}) into Eq. (\ref{E}), one gets for the electric field of the superposition,
\begin{equation}\label{EF}
\begin{split}
\mathbf{E}^{\perp} =\sqrt{\frac{2}{\varepsilon_0}}
& [ \hat{x} (\cos\vartheta \cos{k_\perp x} +\cos{k_\perp y})         \\
& -i\hat{z} \sin\vartheta \sin{k_\perp x}] \exp(i k_\parallel z),
\end{split}
\end{equation}
the transverse component of which is polarized only in the $x$ direction, where $k_\perp =k \sin \vartheta$ and $k_\parallel =k \cos \vartheta$.
According to Maxwell's equations, the corresponding magnetic field assumes
\begin{equation}\label{MF}
\begin{split}
\mathbf{H}^{\perp} =\sqrt{\frac{2}{\mu_0}}
& [ \hat{y} (\cos\vartheta \cos{k_\perp y} +\cos{k_\perp x})        \\
& -i\hat{z} \sin\vartheta \sin{k_\perp y}] \exp(i k_\parallel z).
\end{split}
\end{equation}
Its transverse component is polarized only in the $y$ direction. Both electric field (\ref{EF}) and magnetic field (\ref{MF}) have $z$-polarized longitudinal components.

\begin{figure}[t]
	\centering
	\includegraphics[width=8cm]{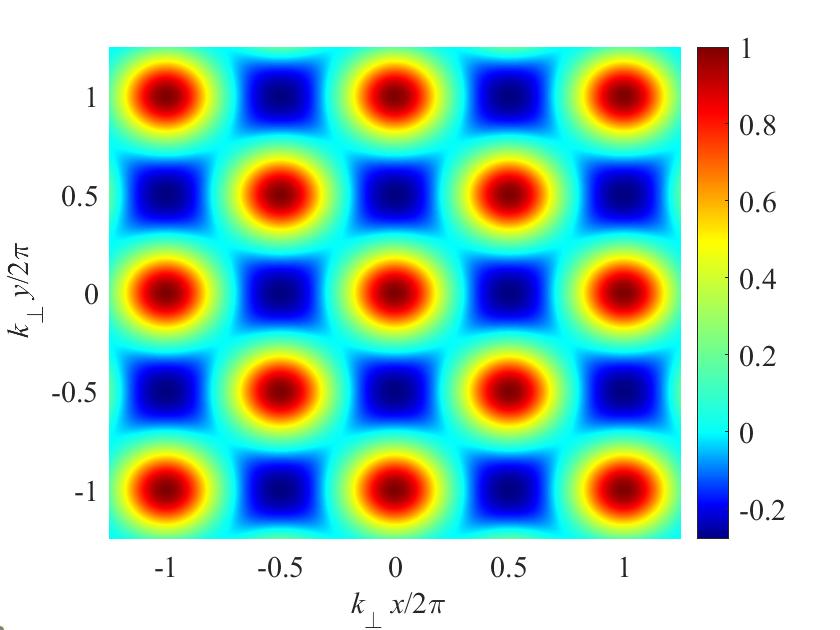}
	\caption{Distribution of normalized $g^\perp_z$ in the transverse plane, where $\vartheta=2 \pi/5$, $|k_\perp x|/2 \pi \le 1.25$, and $|k_\perp y|/2 \pi \le 1.25$.}
	\label{Fig:perp}
\end{figure}

According to expression (\ref{PoyV}) for the Poynting vector, it is straightforward to make use of electric and magnetic fields (\ref{EF})-(\ref{MF}) to obtain the $z$-component of the Poynting vector, which reads
\begin{equation}\label{Z-PV}
\begin{split}
g^\perp_z =& c[\cos\vartheta (\cos{k_\perp x} +\cos{k_\perp y})^2   \\
           & +(1-\cos\vartheta)^2 \cos{k_\perp x} \cos{k_\perp y}],
\end{split}
\end{equation}
where
$c=1/\sqrt{\varepsilon_0 \mu_0}$.
The first term is non-negative. But the second one can be. As a result, there are domains in which $g^\perp_z$ is less than zero. After all, there always exist such points at which one has
$\cos(k_\perp x)=-\cos(k_\perp y) \neq 0$
and therefore $g^\perp_z <0$.
It is seen that the first non-negative term is proportional to $\cos\vartheta$. So under the extremely non-paraxial condition in which
$\vartheta =\pi/2$,
Eq. (\ref{Z-PV}) reduces to
$g^\perp_z =c \cos{kx} \cos{ky}$.
The maximum of its negative value is equal to the maximum of its positive value. In this case, electric field (\ref{EF}) and magnetic field (\ref{MF}) reduce to
\begin{equation*}
\begin{split}
\mathbf{E}^{\perp} =\sqrt{\frac{2}{\varepsilon_0}}(\hat{x} \cos{ky}-i\hat{z} \sin{kx}), \\
\mathbf{H}^{\perp} =\sqrt{\frac{2}{\mu_0}}        (\hat{y} \cos{kx}-i\hat{z} \sin{ky}),
\end{split}
\end{equation*}
respectively. The amplitude of their ``longitudinal'' components is the same as that of their ``transverse'' components.
On the other hand, in the zeroth-order paraxial approximation in which $\sin\vartheta \approx 0$, one has
$g^\perp_z \approx 4c$,
which is positive. This is because in such an approximation, expressions (\ref{EF}) and (\ref{MF}) tend to the electric and magnetic fields of a linearly-polarized plane wave,
\begin{equation*}
\mathbf{E}^{\perp} \approx 2 \sqrt{\frac{2}{\varepsilon_0}} \hat{x} \exp(i k z), \quad
\mathbf{H}^{\perp} \approx 2 \sqrt{\frac{2}{\mu_0}}         \hat{y} \exp(i k z),
\end{equation*}
respectively. Their longitudinal components all vanish. As a matter of fact, in the first-order paraxial approximation in which
$\sin\vartheta \approx \vartheta$ and $\cos\vartheta \approx 1$, one has
$g_z^\perp \approx c (\cos{k \vartheta x}+\cos{k \vartheta y})^2$,
which is non-negative. This indicates that the negative value of $g_z^\perp$ comes from higher-order terms. Therefore, the larger the acute angle $\vartheta$ is, the bigger the maximum of the negative value of $g_z^\perp$ is. Expression (\ref{Z-PV}) can also be written as
\begin{equation*}
\begin{split}
g^\perp_z =& c[\cos\vartheta (\cos{k_\perp x}-\cos{k_\perp y})^2   \\
           & +(1+\cos\vartheta)^2 \cos{k_\perp x} \cos{k_\perp y}],
\end{split}
\end{equation*}
showing that $g^\perp_z \ge 0$ on the bisectrices $x= \pm y$. A typical distribution of normalized $g^\perp_z$ in the transverse plane is illustrated in Fig. \ref{Fig:perp}, where $\vartheta=2 \pi/5$.

It is worthwhile to look at how the $z$-component of the transient Poynting vector varies with the time \cite{Kais}. To this end, we note that the real-valued electric and magnetic fields of the superposition are
\begin{equation*}
\begin{split}
\boldsymbol{\mathcal E}^\perp &=\frac{1}{2} [\mathbf{E}^\perp \exp(-i \omega t)
                                            +\mathbf{E}^{\perp *} \exp(i \omega t)],  \\
\boldsymbol{\mathcal H}^\perp &=\frac{1}{2} [\mathbf{H}^\perp \exp(-i \omega t)
                                 +\mathbf{H}^{\perp *} \exp(i \omega t)],
\end{split}
\end{equation*}
respectively, where $\mathbf{E}^\perp$ and $\mathbf{H}^\perp$ are given by Eqs. (\ref{EF})-(\ref{MF}). According to the definition,
$\boldsymbol{\mathcal G} =\boldsymbol{\mathcal E} \times \boldsymbol{\mathcal H}$,
for the transient Poynting vector, the $z$-component in this field takes the form
\begin{equation*}
{\mathcal G}_z^\perp =2 g_z^\perp \cos^2 (k_\parallel z -\omega t),
\end{equation*}
where $g_z^\perp$ is given by Eq. (\ref{Z-PV}).

\begin{figure}[t]
	\centering
	\includegraphics[width=8cm]{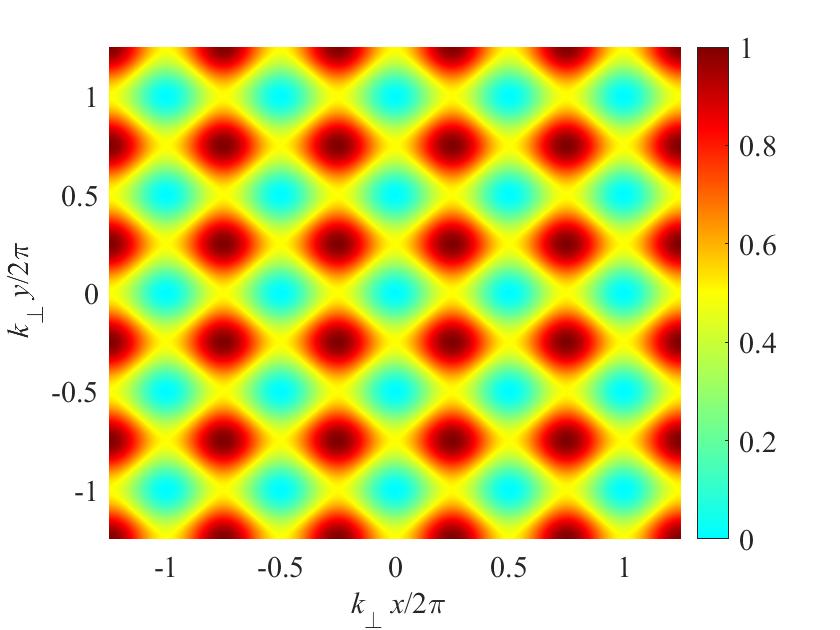}
	\caption{Distribution of normalized $g^{\parallel}_z$ in the transverse plane, where the value of $\vartheta$ is the same as in Fig. \ref{Fig:perp}, $|k_\perp x|/2 \pi \le 1.25$, and $|k_\perp y|/2 \pi \le 1.25$.}
	\label{Fig:para}
\end{figure}

The $g^\perp_z$ in (\ref{Z-PV}) being negative in some domains is similar to what Katsenelenbaum found in Ref. \cite{Kats}. To demonstrate that this phenomenon lies with the specific states of linear polarization expressed by polarization vectors (\ref{PV}), we only adjust the polarization vectors of the four plane waves, changing them into
\begin{equation}\label{PV'}
\begin{split}
\mathbf{a}^\parallel_1 &= \hat{x} \cos \vartheta-\hat{z} \sin \vartheta,  \\
\mathbf{a}^\parallel_2 &= \hat{y} \cos \vartheta-\hat{z} \sin \vartheta,                                   \\
\mathbf{a}^\parallel_3 &=-\hat{x} \cos \vartheta-\hat{z} \sin \vartheta,  \\
\mathbf{a}^\parallel_4 &=-\hat{y} \cos \vartheta-\hat{z} \sin \vartheta,
\end{split}
\end{equation}
where the meaning of the superscript $\parallel$ will be clear in the next section. The same as polarization vectors (\ref{PV}), all the polarization vectors in (\ref{PV'}) stand for states of linear polarization.
In this situation, electric field (\ref{E}) of the superposition becomes
\begin{equation}\label{EF'}
\begin{split}
\mathbf{E}^\parallel = \sqrt{\frac{2}{\varepsilon_0}} i
&[ (\hat{x} \sin{k_\perp x} +\hat{y} \sin{k_\perp y}) \cos\vartheta                  \\
&+ i\hat{z} (\cos{k_\perp x}+ \cos{k_\perp y}) \sin\vartheta ] \exp(i k_\parallel z),
\end{split}
\end{equation}
which has a $y$-polarized transverse component in addition to the $x$-polarized one.
Accordingly, the corresponding magnetic field becomes
\begin{equation}\label{MF'}
\mathbf{H}^{\parallel} =\sqrt{\frac{2}{\mu_0}}
                        i (\hat{y} \sin{k_\perp x} -\hat{x} \sin{k_\perp y}) \exp(i k_\parallel z).
\end{equation}
The same as electric field (\ref{EF'}), it has both $x$- and $y$-polarized transverse components. But its longitudinal component disappears.
As a consequence, the $z$-component of the Poynting vector takes the form,
\begin{equation}\label{Z-PV'}
g^\parallel_z =c \cos\vartheta (\sin^2{k_\perp x} +\sin^2{k_\perp y}).
\end{equation}
In contrast with $g_z^\perp$, $g_z^\parallel$ is non-negative. It vanishes at points
$k_\perp x=m \pi$ and $k_\perp y=n \pi$,
where $m$ and $n$ are integers. It is maximum at points
$k_\perp x=(m+1/2)\pi$ and $k_\perp y=(n+1/2)\pi$.
A typical distribution of normalized $g^{\parallel}_z$ in the transverse plane is illustrated in Fig. \ref{Fig:para}, where the value of $\vartheta$ is the same as in Fig. \ref{Fig:perp}.
It is interesting to note that whether under the extremely non-paraxial condition or under the zeroth-order paraxial condition, $g^\parallel_z$ completely vanishes. In the former case, $\vartheta =\pi/2$, electric field (\ref{EF'}) has only a $z$-polarized component,
\[\mathbf{E}^\parallel = -\sqrt{\frac{2}{\varepsilon_0}} \hat{z} (\cos{kx}+ \cos{k y});\]
and magnetic field (\ref{MF'}) reduces to
\[\mathbf{H}^{\parallel} =\sqrt{\frac{2}{\mu_0}} i(\hat{y} \sin{kx}-\hat{x} \sin{ky}).\]
So certainly the $z$-component of the Poynting vector in such a field is equal to zero by virtue of Eq. (\ref{PoyV}). Whereas in the latter case, $\sin\vartheta=0$, both the electric and magnetic fields vanish.

\section{\label{Explanation} Characterization in terms of Stratton and Jones vectors}

It is seen that even though the constituent plane waves in both situations are linearly polarized, electric fields (\ref{EF}) and (\ref{EF'}) are different in polarization. 
To consider how to characterize such a difference, we note that the polarization vectors of the constituent plane waves in either situation can be expressed in terms of their Jones vectors via Eq. (\ref{a:alpha}). 
Nevertheless, a well known fact is that as the polarization bases, the unit vectors $\mathbf{u}_i$ and $\mathbf{v}_i$ satisfying Eqs. (\ref{triad}) are arbitrary to the extent that a rotation about the associated wavevector $\mathbf{k}_i$ can be performed. By this it is meant that the Jones vector must be defined with respect to some polarization bases.
But on the other hand, Eqs. (\ref{triad}) tell that the polarization bases must depend on the wavevector. Since the wavevectors of the four plane waves are all different, their polarization bases cannot be the same. A consistent way to specify the polarization bases of all the plane waves is to resort to the Stratton vector $\mathbf I$ as is shown by Eqs. (\ref{PB}).
It should be pointed out, however, that for an arbitrarily given superposition field, the Jones vectors of its constituent plane waves defined this way are in general not the same. They are usually different for different plane waves. For instance, one cannot find a common Jones vector for the plane waves in the superposition field that was considered by Katsenelenbaum no matter what Stratton vector is chosen.
But, in principle, the Jones vector in Eq. (\ref{a:alpha}) does not necessarily depend on the wavevector.
Therefore, there must be superposition fields in which the Jones vectors of the constituent plane waves so defined are the same.
These are the non-paraxial fields that we are exclusively concerned with in the present paper. The two superposition fields discussed in previous section were constructed in just this way.

It is easily checked that in either situation, the polarization vectors of the four plane waves can share the same Jones vector
$\alpha=\bigg(\begin{array}{c}
                 1 \\ 0
              \end{array}
        \bigg)$.
But the polarization bases that define this Jones vector are specified by different Stratton vectors. In the first situation, the polarization bases are specified by the Stratton vector $\mathbf{I}^{\perp}=-\hat{x}$ that is perpendicular to the $z$ axis. Whereas in the second situation, the polarization bases are specified by the Stratton vector $\mathbf{I}^{\parallel} =\hat{z}$, which is parallel to the $z$ axis. The difference between the states of polarization of the two superposition fields is thus characterized. From these discussions it follows that when the common Jones vector of the plane waves is fixed, the polarization of superposition field (\ref{E}) depends on the choice of the Stratton vector.

So far we have only chosen one specific common Jones vector
$\alpha=\bigg(\begin{array}{c}
                 1 \\ 0
              \end{array}
        \bigg)$
to illustrate how the Poynting vector reflects the dependence of the polarization of the superposition field on the Stratton vector. Let us further look at how the Poyning vector reflects the dependence of the polarization on the common Jones vector in each situation.
When $\mathbf{I}=\mathbf{I}^\perp$, electric field (\ref{E}) for any particular common Jones vector takes the form
\begin{equation}\label{EF:perp}
\mathbf{E}^\perp (\mathbf{x})=\alpha_1 \mathbf{U}^\perp +\alpha_2 \mathbf{V}^\perp ,
\end{equation}
where
\begin{equation*}
\begin{split}
\mathbf{U}^\perp &=\frac{1}{\sqrt{2 \varepsilon_0}} \sum_{i=1}^{4} \mathbf{u}^\perp_i
                   \exp(i \mathbf{k}_i \cdot \mathbf{x})                                   \\
                 &=\sqrt{\frac{2}{\varepsilon_0}} [\hat{x} (\cos\vartheta \cos{k_\perp x} +\cos{k_\perp y})
                   -i\hat{z}  \sin\vartheta \sin{k_\perp x}] \exp(i k_\parallel z),        \\
\mathbf{V}^\perp &=\frac{1}{\sqrt{2 \varepsilon_0}} \sum_{i=1}^{4} \mathbf{v}^\perp_i
                   \exp(i \mathbf{k}_i \cdot \mathbf{x})                                   \\
                 &=\sqrt{\frac{2}{\varepsilon_0}} [\hat{y} (\cos\vartheta \cos{k_\perp y} +\cos{k_\perp x})
                   -i\hat{z}  \sin\vartheta \sin{k_\perp y}] \exp(i k_\parallel z),
\end{split}
\end{equation*}
and
$\mathbf{u}^\perp_i$ and $\mathbf{v}^\perp_i$ are given by Eqs. (\ref{PB}) with $\mathbf I$ being replaced with $\mathbf{I}^\perp$. 
The corresponding magnetic field assumes
\begin{equation}\label{MF:perp}
   \mathbf{H}^\perp (\mathbf{x})
  =\sqrt{\frac{\varepsilon_0}{\mu_0}} (\alpha_1 \mathbf{V}^\perp -\alpha_2 \mathbf{U}^\perp).
\end{equation}
If
$\alpha=\bigg(\begin{array}{c}
                 1 \\ 0
              \end{array}
        \bigg)$,
electric field (\ref{EF:perp}) and magnetic field (\ref{MF:perp}) go back to (\ref{EF}) and (\ref{MF}), respectively. 
Substituting Eqs. (\ref{EF:perp}) and (\ref{MF:perp}) into Eq. (\ref{PoyV}), one gets for the Poynting vector in this field,
\begin{equation}\label{PV:perp}
\begin{split}
\mathbf{g}^\perp =\hat{z} g^\perp_z
   -\sigma [\hat{x} & (\cos{k_\perp x}+\cos\vartheta \cos{k_\perp y}) \sin{k_\perp y} \\
   -\hat{y} & (\cos{k_\perp y}+\cos\vartheta \cos{k_\perp x}) \sin{k_\perp x}] \sin\vartheta,
\end{split}
\end{equation}
where the constant factor $c$ is omitted, $g^\perp_z$ is given by Eq. (\ref{Z-PV}), and
$\sigma= -i(\alpha^*_1 \alpha_2 -\alpha^*_2 \alpha_1)$
is the ellipticity of polarization.
When $\mathbf{I}=\mathbf{I}^\parallel$, electric field (\ref{E}) for any particular common Jones vector becomes
\begin{equation}\label{EF:para}
\mathbf{E}^\parallel (\mathbf{x})=\alpha_1 \mathbf{U}^\parallel +\alpha_2 \mathbf{V}^\parallel,
\end{equation}
where
\begin{equation*}
\mathbf{U}^\parallel =\frac{1}{\sqrt{2 \varepsilon_0}} \sum_{i=1}^{4} \mathbf{u}^\parallel_i
                                     \exp(i \mathbf{k}_i \cdot \mathbf{x}),                \quad
\mathbf{V}^\parallel =\frac{1}{\sqrt{2 \varepsilon_0}} \sum_{i=1}^{4} \mathbf{v}^\parallel_i
                                     \exp(i \mathbf{k}_i \cdot \mathbf{x}),
\end{equation*}
and
$\mathbf{u}^\parallel_i$ and $\mathbf{v}^\parallel_i$ are given by Eqs. (\ref{PB}) with $\mathbf I$ being replaced with $\mathbf{I}^\parallel$. The same procedure gives for the resultant Poynting vector,
\begin{equation}\label{PV:para}
\begin{split}
\mathbf{g}^\parallel &= \hat{z} g^\parallel_z                            \\
                     &+ \frac{\sigma}{2}
                       (\hat{x} \sin{k_\perp y}-\hat{y} \sin{k_\perp x}) (\cos{k_\perp x}+\cos{k_\perp y}) \sin{2\vartheta},
\end{split}
\end{equation}
where $g^\parallel_z$ is given by Eq. (\ref{Z-PV'}).

Expression (\ref{PV:perp}) or (\ref{PV:para}) shows that given the Stratton vector, the transverse component \cite{Imbe} of the Poynting vector depends on the common Jones vector through $\sigma$ though the $z$-component does not. This reflects the fact that given the Stratton vector, the common Jones vector of the constituent plane waves is able to completely describe the polarization state of superposition field (\ref{EF:perp}) or (\ref{EF:para}). So a comparison of (\ref{EF:perp}) and (\ref{EF:para}) confirms our previous observation that when the common Jones vector of the plane waves is fixed, the polarization state of superposition field (\ref{E}) depends on the choice of the Stratton vector. The difference between expressions (\ref{PV:perp}) and (\ref{PV:para}) just reflects such a dependence.
It is thus concluded that only when combined with the Stratton vector can the common Jones vector of the plane waves be able to completely describe the state of polarization of their superposition field. We are now in a position to have to explain how the common Jones vector of the plane waves combines with the Stratton vector to describe the state of polarization of the superposition field.

\section{\label{Generalization} Generalization of Jones vector to non-paraxial fields}

It is noted that both superposition fields (\ref{EF:perp}) and (\ref{EF:para}) take the form of (\ref{E:UV}). Being the common Jones vector of all the constituent plane waves, $\alpha$ in expression (\ref{E:UV}) could be viewed as the Jones vector of the entire field $\mathbf E$. By this it is meant that the complex-valued vector functions $\mathbf U$ and $\mathbf V$ serve as the polarization bases.
Such polarization bases are different from those for the electric field of a plane wave in that they are dependent on the position $\mathbf x$ and, therefore, usually do not obey
$\mathbf{U}^* \cdot \mathbf{V} =0$
as can be readily checked with $\mathbf{U}^\perp$ and $\mathbf{V}^\perp$ in Eq. (\ref{EF:perp}).
However, with the help of Eqs. (\ref{PB}), it is not difficult to prove that they are orthogonal to each other in the following sense,
\begin{equation}\label{orth}
   \iint \mathbf{U}^* \cdot \mathbf{V} \mathrm{d}x \mathrm{d}y
  =\iint \mathbf{V}^* \cdot \mathbf{U} \mathrm{d}x \mathrm{d}y  =0.
\end{equation}
Moreover, they are equal in ``magnitude'',
 \[\iint \mathbf{U}^* \cdot \mathbf{U} \mathrm{d}x \mathrm{d}y
=  \iint \mathbf{V}^* \cdot \mathbf{V} \mathrm{d}x \mathrm{d}y \equiv W.\]
Combining the polarization bases $\mathbf U$ and $\mathbf V$ into a $3 \times 2$ matrix \cite{Li08},
$\Pi=(\begin{array}{lr}
\mathbf{U} & \mathbf{V}
\end{array})$,
Eq. (\ref{E:UV}) can be expressed in terms of $\alpha$ explicitly as
\begin{equation*}
\mathbf{E}(\mathbf{x}) =\Pi \alpha.
\end{equation*}
Multiplying this equation by $\Pi^\dag$ on the left, taking the integral of both sides over the transverse plane, and considering the above-mentioned properties of $\mathbf U$ and $\mathbf V$, we have
\begin{equation*}
\alpha =\frac{1}{W} \iint \Pi^\dag \mathbf{E} \mathrm{d}x \mathrm{d}y.
\end{equation*}
It indicates that $\alpha$ in expression (\ref{E:UV}) is indeed the Jones vector of the superposition field $\mathbf E$, representing the projections of $\mathbf E$ onto the polarization bases $\Pi$. It is a global quantity. It does not depend on the wavevector $\mathbf{k}_i$ of the constituent plane wave nor on the position $\mathbf x$ in the superposition field. This means that what we have actually done before is to generalize the notion of Jones vector for plane waves to non-paraxial fields. Akin to the Jones vector for a plane wave, the Jones vector for a non-paraxial field must be defined with respect to the polarization bases specified by some Stratton vector via Eqs. (\ref{UV}) and (\ref{PB}). Given the Stratton vector, the Jones vector describes the polarization state of non-paraxial field (\ref{E:UV}). But on the other hand, if the Jones vector is fixed, the polarization state is dependent on the choice of the Stratton vector through the polarization bases. In a word, the Stratton vector shows up as a degree of freedom that combines with the Jones vector to completely describe the state of polarization of non-paraxial fields.

As mentioned above, the polarization bases $\mathbf U$ and $\mathbf V$ for superposition field (\ref{E:UV}) vary with the position $\mathbf x$, in contrast with the polarization bases $\mathbf{u}_i$ and $\mathbf{v}_i$ for the constituent plane wave, which have nothing to do with the position. In particular, the Stratton vector specifies the way in which they depend on the position via Eqs. (\ref{PB}) and (\ref{UV}). So even though the Jones vector $\alpha$ is global, the property described by the Jones vector, the polarization of superposition field (\ref{E:UV}), is not. It varies with the position. The Stratton vector determines how the polarization depends on the position.
If we refer to the position-dependent properties of an electromagnetic field as its locality properties, the polarization is such a property. This may explain why the Poynting vector, another locality property of the electromagnetic field, can aptly reflect the different roles of the Stratton and Jones vectors in the polarization. Indeed, substituting electric field (\ref{E:UV}) and its corresponding magnetic field into Eq. (\ref{PoyV}), we find
\begin{equation}\label{g:sigma}
   \mathbf{g}=\frac{\varepsilon_0}{2}\Big[\Re(\mathbf{U} \times \mathbf{V}^*)
             +\frac{i}{2}\sigma(\mathbf{U} \times \mathbf{U}^* +\mathbf{V} \times \mathbf{V}^*)\Big],
\end{equation}
where the constant factor $c$ is omitted as before. The first term does not depend on the Jones vector. Only the second term does. It depends merely on the choice of the Stratton vector via the polarization bases.
Because this is the term that is responsible for the $z$-component of Poynting vectors (\ref{PV:perp}) and (\ref{PV:para}), the negative value of the $z$-component of Poynting vector (\ref{PV:perp}) in some domains embodies the concrete locality property of the polarization bases $\mathbf{U}^\perp$ and $\mathbf{V}^\perp$ specified by the specific Stratton vector $\mathbf{I}^\perp =-\hat{x}$.

We stress that the conclusions drawn here can readily be extended to a non-paraxial monochromatic field of  continuously-distributed angular spectrum. Specifically, the polarization bases $\mathbf U$ and $\mathbf V$ that define the Jones vector $\alpha$ in Eq. (\ref{E:UV}) can take the form
\begin{equation*}
\begin{split}
\mathbf{U}= & \frac{1}{\sqrt{2 \varepsilon_0}} \iint \frac{e(\mathbf{k})}{2 \pi}
              \mathbf{u}(\mathbf{k}) \exp(i \mathbf{k} \cdot \mathbf{x}) \mathrm{d}k_x \mathrm{d}k_y, \\
\mathbf{V}= & \frac{1}{\sqrt{2 \varepsilon_0}} \iint \frac{e(\mathbf{k})}{2 \pi}
              \mathbf{v}(\mathbf{k}) \exp(i \mathbf{k} \cdot \mathbf{x}) \mathrm{d}k_x \mathrm{d}k_y,
\end{split}
\end{equation*}
where $e(\mathbf{k})$ is the scalar angular spectrum, unit-vector functions $\mathbf{u}(\mathbf{k})$ and $\mathbf{v}(\mathbf{k})$ denote the polarization bases of the constituent plane wave, which are specified by the Stratton vector as follows,
\begin{equation}\label{SV}
\mathbf{u} =\mathbf{v} \times \frac{\mathbf k}{k},                                \quad
\mathbf{v} =\frac{\mathbf{I} \times \mathbf{k}} {|\mathbf{I} \times \mathbf{k} |}.
\end{equation}
Eqs. (\ref{SV}) guarantee that they satisfy the orthogonal property (\ref{orth}). Moreover, they are equal in ``magnitude'',
 \[\iint \mathbf{U}^* \cdot \mathbf{U} \mathrm{d}x \mathrm{d}y
=  \iint \mathbf{V}^* \cdot \mathbf{V} \mathrm{d}x \mathrm{d}y
=\frac{1}{2 \varepsilon_0} \iint |e(\mathbf{k})|^2  \mathrm{d}k_x \mathrm{d}k_y . \]
In particular, expression (\ref{g:sigma}) for the Poynting vector is valid for the resultant field.

\section{\label{Conclusions} Conclusions and remarks}

In summary, we succeeded in generalizing the notion of Jones vector of plane waves to non-paraxial fields through investigating the dependence of the Poynting vector on the state of polarization in the non-paraxial superposition of four plane waves. We found that the generalized Jones vector $\alpha$ in expression (\ref{E:UV}) is not able to completely describe the polarization of the non-paraxial field $\mathbf E$.
In addition to the Jones vector, the Stratton vector $\mathbf I$, which specifies the locality property of the polarization bases through Eqs. (\ref{PB}) and (\ref{UV}), turns out to be another degree of freedom. It determines how the polarization described by the Jones vector depends on the position. The different roles of the Stratton and Jones vectors in the local polarization are well reflected in the Poynting vector (\ref{g:sigma}).
We showed that when the Jones vector is fixed, the Poynting vector depends on the choice of the Stratton vector. On the other hand, when the Stratton vector is chosen to be either $\mathbf{I}^\perp$ or $\mathbf{I}^\parallel$, the transverse component of the Poynting vector depends on the Jones vector though the $z$-component does not.

It is worth noting that because of the linearity of Maxwell's equations, one is allowed to consider a linear combination of two non-paraxial fields that share the same Stratton vector but have different Jones vectors.
The situation discussed by Katsenelenbaum in Ref. \cite{Kats} is just one such example.
Letting Katsenelenbaum's electric field be expressed by Eq. (\ref{E}), the polarization vectors of its constituent plane waves assume
\begin{equation*}
\begin{split}
\mathbf{a}_1 &= \hat{y},                                         \\
\mathbf{a}_2 &=-\hat{y} \cos \vartheta +\hat{z} \sin \vartheta,  \\
\mathbf{a}_3 &=-\hat{y},                                         \\
\mathbf{a}_4 &= \hat{y} \cos \vartheta +\hat{z} \sin \vartheta.
\end{split}
\end{equation*}
(The minus sign in front of $\hat y$ in $\mathbf{a}_2$ was omitted in Ref. \cite{Kats}.)
They can be divided into two parts. One is made of $\mathbf{a}_1$ and $\mathbf{a}_3$. Another is made of $\mathbf{a}_2$ and $\mathbf{a}_4$. The polarization bases in both parts are specified by the same Stratton vector $\mathbf{I}^\parallel =\hat{z}$. But as can be readily checked, the Jones vector in the former is
$\bigg(\begin{array}{c}
          0 \\ 1
       \end{array}
 \bigg)$
and the Jones vector in the latter is
$\bigg(\begin{array}{c}
          -1 \\ 0
       \end{array}
 \bigg)$.
Of course, for the same reason, one is also allowed to consider a linear combination of two non-paraxial fields that share the same Jones vector but have different Stratton vectors. Further discussions are beyond the scope of the present paper.
It is hoped that a deep understanding of the role of the Stratton vector in the description of the polarization of non-paraxial fields will help to resolve the controversy \cite{Kats,Kotl-SNKP,Novi-N} over the occurrence of the local reverse energy flow.

%\end{acknowledgments}


%\bibliography{reference}

%\bibliographyfullrefs{reference}

\end{document}